\documentclass[aps, prd, tightenlines, floatfix, onecolumn, amsmath, superscriptaddress]{revtex4}
\usepackage{graphicx}
\usepackage{dcolumn}
\usepackage{bm}
\usepackage{float}
\usepackage{hyperref}
\usepackage{bigints}
\usepackage{xcolor}

\begin{document}

\title{Assessment of ALP scenarios for GRB 221009A}

\author{Giorgio Galanti}
\email{gam.galanti@gmail.com}
\affiliation{INAF, Istituto di Astrofisica Spaziale e Fisica Cosmica di Milano, Via Alfonso Corti 12, I -- 20133 Milano, Italy}

\author{Marco Roncadelli}
\email{marcoroncadelli@gmail.com}
\affiliation{INFN, Sezione di Pavia, Via A. Bassi 6, I -- 27100 Pavia, Italy}

\author{Fabrizio Tavecchio}
\email{fabrizio.tavecchio@inaf.it}
\affiliation{INAF, Osservatorio Astronomico di Brera, Via E. Bianchi 46, I -- 23807 Merate, Italy}

\date{\today}
\begin{abstract}
About one month after the revolutionary discovery of the Gamma Ray Burst (GRB) GRB 221009A and intense theoretical efforts to explain its detection, time seems to us ripe to make an assessment of the axion-like particle (ALP) based scenarios, since it is a common belief that conventional physics would have prevented such a detection. We overcome the almost complete lack of information -- so far only astronomical telegrams have been released -- by relying as much as possible upon the analogy with the emission from the GRB 190114C detected by the MAGIC collaboration in 2019, since it was the highest energy GRB detected before and for a time lapse similar to that over which GRB 221009A has been observed. 
\end{abstract}



\maketitle



\section{Introduction}

The revolutionary discovery of GRB 221009A at redshift $z = 0.151$~\cite{32648} at energy up to ${\cal E} = 18 \, {\rm TeV}$ made by the LHAASO collaboration lasting $2000 \, {\rm s}$ after the trigger $t_0$~\cite{lhaaso} and at ${\cal E} = 251 \, {\rm TeV}$ by the Carpet-2 collaboration at $4536 \, {\rm s}$ after $t_0$~\cite{carpet} strongly challenges conventional physics. The trigger $t_0$ has been set by {\it Fermi}/GBM on October 10, 2022 at 13:16:60 UT~\cite{atFermi32658}. Throughout this paper we suppose that both observations indeed concern GRB 221009A. 

Basically, at such very-high-energies (VHE, ${\cal E} > 100 \, {\rm GeV}$) photons from GRB 221009A scatter off photons of the Extragalactic Background Light (EBL) -- the diffuse light emitted by all stars during the cosmic evolution -- thereby producing $e^+ e^-$ pairs, and so largely depleting the photon beam (for a review, see~\cite{dwek}). Because axion-like particles (ALPs) reduce such a dimming through photon-ALP conversions and 
reconversions in the presence of a magnetic field -- as first recognized in~\cite{drm2007} -- several papers have considered ALP scenarios with different properties~\cite{grtClu,bhm,troitskyGRB,messicani,carenzamarsh}. 

Very recently, Carenza and Marsh made an overview of them (thereafter CM)~\cite{carenzamarsh}. Our goal is very similar in nature to that of such an overview, but it greatly differers in its content. Given the almost complete lack of information -- only astronomical telegrams have been released so far -- we try to use as much as possible what has been learnt from GRB 190114C, the previous highest energy GRB detected by the MAGIC collaboration in 2019, which has been observed up to an energy ${\cal E} \simeq 1 \, {\rm TeV}$~\cite{magic1,magic2}. We prefer to consider GRB 190114C since it has been observed during the time interval $62 \, {\rm s} \lesssim t \lesssim 2454 \, {\rm s}$ after the trigger -- which is of the same order of magnitude of the time interval over which GRB 221009A has been detected -- while GRB 190929A reaching an energy ${\cal E} \simeq 3.3 \, {\rm TeV}$ has been monitored by the H.E.S.S. collaboration during the time lapse $1.5 \cdot 10^4 \, {\rm s} \lesssim t \lesssim 2 \cdot 10^5 \, {\rm s}$ after the trigger~\cite{hess} -- much larger than that of GRB 221009A -- hence less similar to GRB 221009A (see also~\cite{nava2021}). So, throughout this paper GRB 190114C will be considered as a sort of `smaller twin' of GRB 221009A as far as the emission is concerned.

\section{Some preliminaries}

Let us first recall a few basic properties of the GRBs  which will be useful later. First of all, short GRBs (with the prompt emission lasting less than 2 seconds) originate from the merging of two neutron stars (or a neutron star with a black hole), whereas long GRBs (with the prompt emission lasting more than 2 seconds) arise from ultra-relativistic jets launched from the collapsing cores of dying massive stars. Moreover, GRBs are characterized by two phases: an initial {\it prompt} and a subsequent {\it afterglow}. The prompt emission comes from the jet, is highly variable and has its maximum in the keV-MeV band. Its duration lasts from milliseconds to minutes. Thereafter, the afterglow emission takes over -- also rapidly varying -- which is due to the shock waves produced by the interaction of the jet with the external medium and can last up to months. The highest frequencies are generated at the beginning of the afterglow, and as time goes by progressively lower and lower frequencies are emitted, spanning the whole electromagnetic spectrum from the gamma band down to the radio band. According to this view, GRB 221009A has been observed at the beginning of the afterglow, and not in the prompt emission as stated in~\cite{bhm}. While it is generally believed that the prompt emission is explained as synchrotron radiation by electrons accelerated in the jet, the spectral energy distribution (SED) of the afterglow is not so simple and will be discussed later. We emphasize that this is the conventional view based on the fireball model. Nevertheless, alternative models have been proposed, which are however unimportant for our needs. An authoritative review of GRBs up to 2005 is contained in~\cite{piranreview}, while an up-to-date account can be found in~\cite{nava2021}.

\section{Motivation for ALP scenarios}

As we said, the EBL strongly suppresses the flux emitted by GRB 221009A. In order to address this point in a quantitative fashion, we start by recalling that the photon survival probability within conventional physics (CP) is given by
\begin{equation}
P_{\rm CP} ({\cal E}; \gamma \to \gamma) = e^{- \, \tau_{\rm CP} (\cal E)}~,
\label{9112022a}
\end{equation}
where $\tau_{\rm CP} ({\cal E})$ is the optical depth. In the literature several estimates of 
$\tau_{\rm CP} ({\cal E})$ can be found, and a fairly complete list has been reported in~\cite{bhm}. Since we have to make a choice, here we adopt the rather conservative estimate of Franceschini and Rodighiero~\cite{franceschinirodighiero}, which gives $\tau^{\rm FR}_{\rm CP} \simeq 14$ at ${\cal E} = 18 \, {\rm TeV}$ and $\tau^{\rm FR}_{\rm CP} \simeq 15000$ at ${\cal E} = 251 \, {\rm TeV}$. Therefore, the corresponding photon survival probabilities read 
\begin{equation}
P^{\rm FR}_{\rm CP} ({\cal E} = 18 \, {\rm TeV}; \gamma \to \gamma) \simeq 8.5 \times 
10^{- 7}~,
\label{9112022b}
\end{equation}
and 
\begin{equation}
P^{\rm FR}_{\rm CP} ({\cal E} = 251 \, {\rm TeV}; \gamma \to \gamma) = \simeq e^{- 15000}~.
\label{9112022c}
\end{equation}
This can be compared with the model of Gilmore {\it et al.}~\cite{gilmore}, which yields $\tau^{\rm G}_{\rm CP} \simeq 14$ at ${\cal E} = 18 \, {\rm TeV}$ -- identical to the previous case -- while $\tau^{\rm G}_{\rm CP} \simeq 9500$ at ${\cal E} = 251 \, {\rm TeV}$. Accordingly, Eq.~(\ref{9112022b}) remains unchanged but Eq.~(\ref{9112022c}) becomes
\begin{equation}
P^{\rm G}_{\rm CP} ({\cal E} = 251 \, {\rm TeV}; \gamma \to \gamma) \simeq e^{- 9500}~.
\label{9112022d}
\end{equation}
According to the common wisdom, within conventional physics it is impossible to detect photons with these survival probabilities (more about this, in the last Section). This is also emphasized in~\cite{carpet}. 

\begin{figure}[h]
\begin{center}
\includegraphics[width=0.25\textwidth]{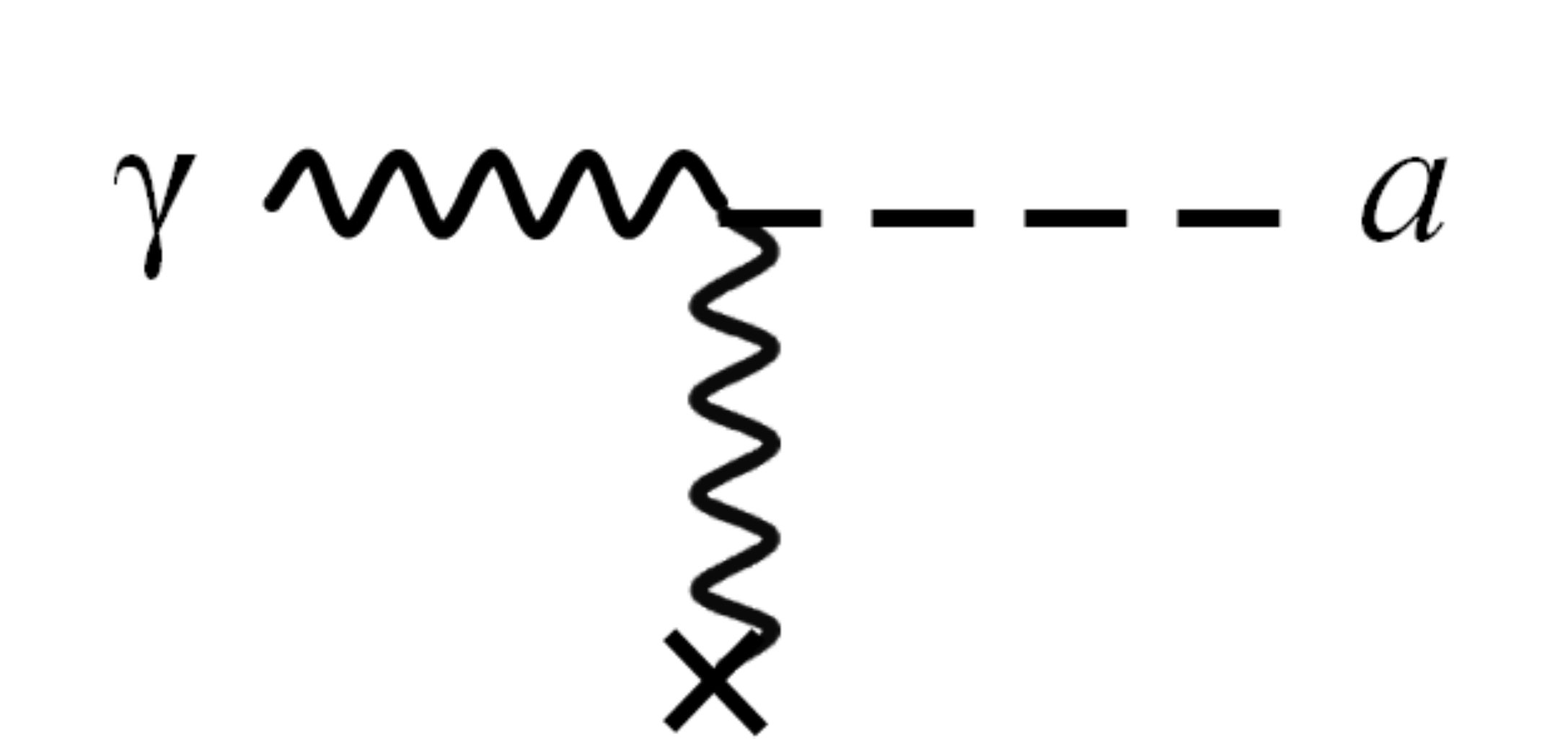}
\caption{\label{immagine1} ALP-two photons vertex with coupling constant $g_{a \gamma \gamma}$.}
\end{center}
\end{figure}

We refrain from discussing the formalism of photon-ALP conversions and reconversions  both because we do not use it here and because it is described in so many papers (for the  motivation of ALPs see~\cite{ringwald1,ringwald2}, and for a review see~\cite{universe} and references therein). The photon-ALP Lagrangian is
\begin{equation}
{\cal L}_{a \gamma} = - \, \frac{1}{4} \, g_{a \gamma \gamma} \, F_{\mu \nu} \, {\tilde{F}}^{\mu \nu} \, a = g_{a \gamma \gamma} \, {\bf E} \cdot {\bf B} \, a~,
\label{9112022e}
\end{equation}
\noindent where $a$ is the ALP field and -- in the present context -- ${\bf E}$ represents the electric field of a propagating photon whereas ${\bf B}$ stands for the external magnetic field. The last term in Eq.~(\ref{9112022e}) is represented by the Feynman diagram in Fig.~\ref{immagine1} 

\noindent where the horizontal photon leg corresponds to ${\bf E}$ while the vertical one to ${\bf B}$. This diagram represents a photon-to-ALP or an ALP-to-photon {\it conversion}, which takes place in principle within any magnetized astronomical object. We stress that -- according to Eq.~(\ref{9112022e}) -- ALPs do not couple either to {\it single} photons or to matter, as explicitly shown in~\cite{grjhea}. 

Thus, a way to reduce the EBL absorption is to have photon-to-ALP conversions in the source or close to it, and ALP-to-photon conversions in the Milky Way since the EBL is fully transparent to the ALPs.  

\begin{figure}[h]
\begin{center}
\includegraphics[width=.50\textwidth]{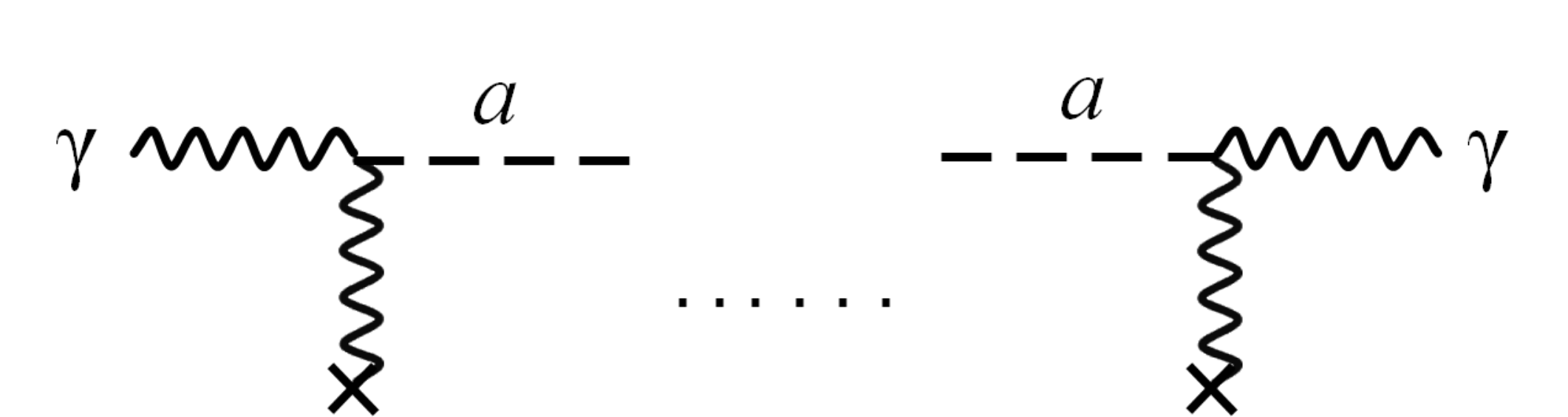}
\caption{\label{fey1.pdf} Schematic view of photon-ALP oscillations.}
\end{center}
\end{figure}

Another strategy to reduce the EBL absorption is based on photon-ALP {\it oscillations}~\cite{mpz,rs} occurring as the photon-ALP beam propagates in extragalactic space, provided that the extragalactic magnetic field is sufficiently strong. Pictorially, this process arises by joining two vertices in Fig.~\ref{immagine1} by one ${\bf E}$ or one ALP horizontal  leg, while the other vertical photon legs represent ${\bf B}$. This is shown in Fig.~\ref{fey1.pdf} 

As a consequence, photons acquire a `split personality': when they behave as true photons they interact with the EBL but when they behave as ALPs they do not. This fact reduces the effective optical depth 
$\tau_{\rm ALP} ({\cal E})$ (accounting for both CP and ALPs). So, Eq.~(\ref{9112022a}) gets now replaced by 
\begin{equation}
P_{\rm ALP} ({\cal E}; \gamma \to \gamma) = e^{- \, \tau_{\rm ALP}(\cal E)}~.
\label{9112022f}
\end{equation}
The crux of the argument is the negative exponential dependence of $P_{\rm ALP} ({\cal E}; \gamma \to \gamma)$ on $\tau_{\rm ALP} ({\cal E})$, since a small reduction of 
$\tau_{\rm ALP} ({\cal E})$ with respect to $\tau_{\rm CP} ({\cal E})$ gives rise to a large enhancement of $P_{\rm ALP} ({\cal E}; \gamma \to \gamma)$ as compared with $P_{\rm CP} ({\cal E}; \gamma \to \gamma)$.

Manifestly, both possibilities can occur at the same time.

\section{Photon-ALP interconversions}

Different authors have chosen different places for the initial photon-to-ALP conversion. Below, we review the various possibilities, adding some remarks. All ALP scenarios for GRB 221009A considered in this paper involve the ALP reconversion to photons in the Milky Way,   basically employing the magnetic field model developed by Jansson and 
Farrar~\cite{jansonfarrar1,jansonfarrar2,BMWturb}. 

\

\noindent {\it Photon-to-ALP conversion in the jet} -- Recalling that GRB 221009A is observed in the early phase of the afterglow, a possibility for such a conversion is inside the jet in such a situation. This scenario has been put forward in the first attempt to understand the GRB 221009A emission~\cite{grtClu}. The job is to fix the free parameters. The ALP 
mass $m_{\rm ALP}$ is taken to be $m_{\rm ALP} \simeq 10^{- 10} \, {\rm eV}$, in line with the previous work of the authors, and the two-photon coupling is chosen as $g_{a \gamma \gamma} \simeq 5 \cdot 10^{- 12} \, {\rm GeV}^{- 1}$ so as to meet the strongest upper bound coming from the magnetized white dwarfs, which reads $g_{a \gamma \gamma} \lesssim 5.4 \cdot 10^{- 12} \, {\rm GeV}^{- 1}$ at the 2$\sigma$ level~\cite{mwd}. Three free parameters are the magnetic field strength in the jet $B_{\rm jet}$, the electron number density in the jet $n_{{\rm jet}, e}$, both when it is at the beginning of the afterglow (which means outside the source but close to it) and the bulk Lorentz factor $\Gamma$. Only an `educated guess' allows to fix them. And such an `educated guess' comes from the analogy with the `smaller twin' GRB 190114C~\cite{magic1,magic2}. In this way the author's choice is $B_{\rm jet} \simeq 1 \, {\rm G}$, $n_{{\rm jet}, e} \simeq 10^3 \, {\rm cm}^{- 3}$ and $\Gamma \simeq 1000$. It goes without saying that fixing the astrophysical parameters is very uncertain business, but nowadays no more reliable estimate is possible. A similar option is chosen also in~\cite{messicani}, where it is instead assumed that $B_{\rm jet} \simeq 10^6 \, {\rm G}$. It is easy to see that such a strong magnetic field makes the photon-to-ALP conversions totally negligible, owing to the one-loop vacuum polarization effect~\cite{rs}.

\ 

\noindent {\it Photon-to-ALP conversion in the host galaxy} -- Other authors prefer to assume that the initial photon-to-ALP conversion occurs only in the host galaxy of GRB 221009A~\cite{bhm,troitskyGRB,messicani,carenzamarsh}, while~\cite{grtClu} considers  also this option. Unfortunately, at the time of writing no information about the host galaxy is available. What is clear is that GRBs are hosted by star forming galaxies, but no agreement among the experts exists as to whether these are normal spiral galaxies or starburst galaxies~\cite{piranreview}. As far as normal spiral galaxies are concerned, the most likely possibility is that a GRB forms in the spiral arms, where star formation is highest. Note that GRB 190114C is hosted by a normal spiral in its centre, but this is an exception and not the rule. Moreover, even if we rely as much as possible on the analogy with GRB 190114C for its emission, this does not imply that the analogy should be true for the host galaxy and for the location of the GRB inside it. Specifically,~\cite{bhm} implicitly considers a spiral galaxy with $B = 0.5 \, {\mu}G$ with coherence length $L_{\rm coh} = 10 \, {\rm Mpc}$,~\cite{troitskyGRB} supposes that the host galaxy is similar to the Milky Way and assumes that the maximal mixing occurs in the host, namely that the photon-to-ALP conversion probability is $1/3$,~\cite{messicani} take the host just equal to the Milky Way, and the same assumption is made in~\cite{carenzamarsh} but supposing that GRB 221009A is located in its centre. 

\

\noindent {\it Photon-to-ALP conversion in extragalactic space} -- Our knowledge of the extragalactic magnetic ${\bf B}_{\rm ext}$ field is still very poor. Observations only tell that its strength is constrained inside the range $10^{- 7} \, {\rm nG} \lesssim B_{\rm ext} \lesssim 1.7 \, {\rm nG}$ on the scale ${\cal O} (1) \, {\rm Mpc}$~\cite{neronov,durrer,pshirkov,podlesnyi}. Nonetheless, since about twenty years it has become customary to described ${\bf B}_{\rm ext}$ by means of a very specific model. It consists of a domain-like network, in which ${\bf B}_{\rm ext}$ is supposed to be homogeneous over a whole domain of size $L_{\rm dom}$ equal to its coherence length, with ${\bf B}_{\rm ext}$ changing randomly its direction from one domain to the next, keeping approximately the same strength. As a consequence, the photon-ALP beam propagation becomes a {\it random process}, and only a single realization at once can be observed. In addition, it is assumed that such a change of direction is abrupt, because then the beam propagation equation is easy to solve~\cite{kronberg1994,grassorubinstein2001}. Such a scenario -- called {\it domain-like sharp-edges} (DLSHE) -- rests on outflows from primeval galaxies, further amplified by turbulence~\cite{reessetti1968,hoyle1969,kronbergleschhopp1999,furlanettoloeb2001}. Common benchmark values are $B_{\rm ext} = {\cal O} (1) \, {\rm nG}$ on a coherence length ${\cal O} (1) \, {\rm Mpc}$, whence $L_{\rm dom} = {\cal O} (1) \, {\rm Mpc}$ (for more details, see~\cite{galantironcadelli20118prd}). In order to be definite, the authors of~\cite{grtClu} choose $B_{\rm ext} = 1 \, {\rm nG}$ and $L_{\rm dom}$ in the range $(0.2-10) \, \rm Mpc$ and with $\langle L_{\rm dom} \rangle = 2 \, \rm Mpc$. But the abrupt change in direction at the interface between two adjacent domains leads to a failure of the DLSHE model at the energies considered here, owing to photon dispersion on the CMB~\cite{raffelteffect}. A way out of this difficulty is to smooth out the sharp edges of the domains, so that the components of ${\bf B}_{\rm ext}$ change continuously across the interface, thereby leading to the {\it domain-like smooth-edges} (DLSME) model, built up in~\cite{galantironcadelli20118prd,kartavtsev}. Only the ALP scenario described in~\cite{grtClu} contemplates photon-ALP oscillations in extragalactic space within the DLSME model. As discussed in detail in~\cite{grjhea}, above a few TeV photon dispersion on the CMB makes the probability for photon-ALP oscillations vanishingly small. Surprisingly, this point has been misunderstood in~\cite{carenzamarsh}, which states: `{\it the extragalactic scenario may explain the LHAASO observation of an 18 TeV photon, but is incapable of explaining the Carpet-2 event. At very high energies, the extragalactic conversion is efficient across the full range of coupling constants that we consider. As explained in}~\cite{troitskyGRB}, {\it ALPs are then continuously reconverted to photons and absorbed over relatively short distances, thereby depleting the photon flux}'. Such a statement is obviously incorrect, since at the energy of Carpet-2 -- namely ${\cal E} = 251 \, {\rm TeV}$ -- no photon-ALP oscillation is possible in extragalactic space, as explained above. 

\section{Spectral energy distribution (SED)}

This is a crucial issue. Until the LHAASO and Carpet-2 collaborations will release their  results the observed SED is unknown. Nevertheless some totally unjustified statements have been made, as we are going to discuss below. 

We recall that the flux $F ({\cal E},t) \equiv d N (t)/(d t d A d {\cal E})$ is related to the SED 
$\nu F_{\nu} ({\cal E},t)$ as
\begin{equation}
\nu F_{\nu} ({\cal E},t) = {\cal E}^2 \, F ({\cal E},t)~.
\label{10112022b}
\end{equation} 

Now, {\it Fermi}/LAT has observed a photon flux in the energy range $1 \, {\rm MeV} \lesssim {\cal E} \lesssim 1 \, {\rm GeV}$ equal to $(6.2 \pm 0.4) \cdot 10^{- 3} \, \gamma \, 
{\rm cm}^{- 2} \, {\rm s}^{- 1}$ in the time interval $200 \, {\rm s} \lesssim t \lesssim 800 \, {\rm s}$ after $t_0$, with a spectral index $\Gamma = - 1.87 \pm 0.04$, and in addition a single photon of energy ${\cal E} = 99.3 \, {\rm GeV}$ at $240 \, {\rm s}$ after $t_0$~\cite{atFermi32658}. Explicitly, the {\it Fermi/LAT} flux can be restated as
\begin{equation}
\int_{100 \, {\rm MeV}}^{1 \, {\rm GeV}} d {\cal E} ~ F ({\cal E},t) = \bigl(6.2 \pm 0.4 \bigr) \cdot 10^{- 3} \, \gamma \, {\rm cm}^{- 2} \, {\rm s}^{- 1}~,  \ \ \ \ \ \ \ \ \ \ \ \ \ 200 \, {\rm s} + t_0 < t < 800 \, {\rm s} + t_0~,
\label{15112022a}
\end{equation} 
which indeed shows that the {\it Fermi/LAT} flux is time-independent over the range $200 \, {\rm s} + t_0 < t < 800 \, {\rm s} + t_0$. The further statement that the spectral index is 
$\Gamma = - 1.87 \pm 0.04$ implies that $F ({\cal E})$ is a power law. Using this information and inserting $F ({\cal E}) \propto {\cal E}^{ - 1.87 \pm 0.04}$ into Eq. (\ref{15112022a}) the normalization constant gets fixed. This has been done in~\cite{bhm}, finding
\begin{equation}
F ({\cal E}) = 2.1 \cdot 10^{- 6} \left(\frac{{\cal E}}{{\rm TeV}} \right)^{- 1.87} \, {\rm cm}^{- 2} \, {\rm s}^{- 1} \, {\rm TeV}^{- 1}~,  \ \ \ \ \ \ \ \ \ \ \ \ \ 200 \, {\rm s} + t_0 < t < 800 \, {\rm s} + t_0~. 
\label{10112022a}
\end{equation}

\begin{figure}[h]
\begin{center}
\includegraphics[width=0.50\textwidth]{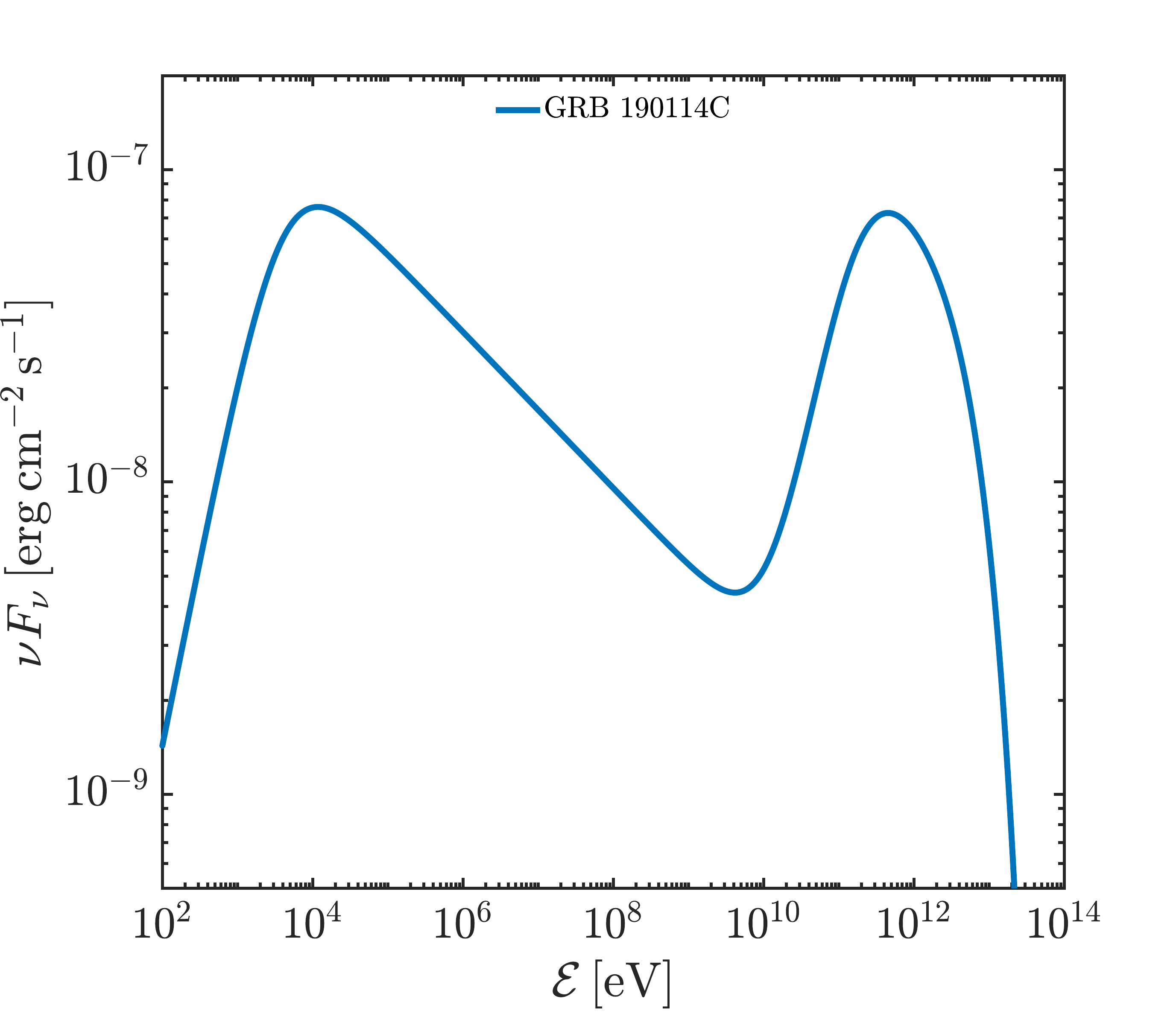}
\caption{\label{figyp} Observed SED of GRB 190114C drawn by us using the parameters reported in~\cite{yamasakipiran}.}
\end{center}
\end{figure}

Both~\cite{bhm} and~\cite{carenzamarsh} have extrapolated the flux in Eq. (\ref{10112022a}) {\it to arbitrarily large energies}. We stress that such an extrapolation up to LHAASO photons implies that the LHAASO flux is {\it time-independent over its full observation time}, namely over $2000 \, {\rm s}$ after $t_0$. We fully disagree for two reasons. First, it looks unreasonable to us that the LHAASO emission remains constant over such a long time as $2000 \, {\rm s}$, since all GRBs observed so far exhibit a very fast variability at any observed frequency. Second, because such extrapolation has no physical motivation. In order to clarify the last point, we still rely upon GRB 190114C, whose observed SED has been re-evaluated by us using the parameters reported in~\cite{yamasakipiran} and is exhibited in Fig.~\ref{figyp}. The reader should keep in mind that the emission varies rapidly in time -- hence the same is true for the observed SED -- as it is evident from the figures in~\cite{yamasakipiran}. Therefore, Fig.~\ref{figyp} is true at a single observation time only.

Because GRB 221009A extends to considerably higher energies with respect to GRB 190114C {\it we do expect that also the inverse Compton peak has been observed} and that it is partially responsible for the emission detected by LHAASO and responsible for the photon observed by Carpet-2. This is the main message we want to convey. Of course, the analogy between the observed SED of GRB 221009A and of GRB 190114C is only qualitative, since the former extends up to $251 \, {\rm TeV}$ while the latter extends only up to about $1 \, {\rm TeV}$. So, for GRB 221009A the inverse Compton peak should be shifted to considerably higher energies, at least during the observation times of LHAASO and Carpet-2. Moreover, it is plausible to have a hardening of the flux in Eq. (\ref{10112022a}) before the peak.

\section{Discussion and conclusions}

We have critically reviewed a few scenarios involving axion-like particles devised to explain the observation of the GRB 221009A. Our conclusions differ in some cases from the similar review paper by Carenza and Marsh~\cite{carenzamarsh}. The novelty of the present paper is that we make up for the lack of information by regarding GRB 190114C as a sort of `smaller twin' of GRB 221009A as far as the emission is concerned. Below, we cursorily discuss the key-features of the various ALP scenarios from our point of view. 

\ 

All authors assume that the ALP-to-photon conversion occurs in the Milky Way. The difference among the various scenarios is threefold.

\begin{itemize}	

\item The site where the photon-to-ALP conversion takes place.

\item Whether photon-ALP oscillations in extragalactic space are considered. 

\item Whether or not a SED (or flux) is assumed.

\end{itemize}

\

Schematically, the situation is as follows.

\

\noindent \cite{grtClu} contemplates the photon-to-ALP conversion both inside the GRB jet at the beginning of the afterglow and in the host galaxy. Also photon-ALP oscillations in extragalactic space are included, but they disappear for energies above a few TeV. Only the photon survival probability is considered -- fluxes are not taken into account -- and so it can explain both the LHAASO and the Carpet-2 results depending on the rapidly time-varying  SED of GRB 221009A. Finally, the assumed value of $g_{a \gamma \gamma}$ is in agreement with the upper bound from the magnetized white dwarfs~\cite{mwd}. Thus, this model is viable. 

\

\noindent \cite{bhm} uses the host galaxy to perform the photon-to-ALP conversion for 
GRB 221009A considered in the propt phase. The power-law flux observed by {\it Fermi/LAT} for $100 \, {\rm MeV} \lesssim {\cal E} \lesssim 1 \, {\rm GeV}$ in Eq. (\ref{10112022a}) is extrapolated to the highest considered energies, contrary to the expectation from GRB 190114C. The assumed value of $g_{a \gamma \gamma}$ depends on the assumed extrapolation and is in disagreement with the upper bound from the magnetized white dwarfs~\cite{mwd}, thereby implying that such a model is not viable.  

\

\noindent \cite{troitskyGRB} also this paper supposes that the photo-to-ALP conversion occurs in the host galaxy. It is {\it assumed} that the maximal mixing occurs in the host -- namely that the photon-to-ALP conversion probability is $1/3$ -- and a close to maximal mixing takes place in the Milky Way. However, this is merely a wishful thinking since one should {\it demonstrate} that a realistic magnetic field in the host and in the Milky Way indeed produces the maximal mixing. And the magnetic field model of Jansson and Farrar~\cite{jansonfarrar1,jansonfarrar2,BMWturb} fails to do the job, so this model is at this stage not viable.

\

\noindent \cite{messicani} considers several topics -- including the leptonic inverse Compton 
scattering -- and among them also an ALP scenario. The photon-to-ALP conversion is supposed to take place both in the jet and in the host galaxy, assumed to have both a morphology and a magnetic field similar to those of the Milky Way. But we already pointed out that no photon-to-ALP conversion can actually occur in the jet with their assumed parameters. Three $(m_{\rm ALP}, g_{a \gamma \gamma})$ pairs of values are chosen, but the value of $g_{a \gamma \gamma}$ always exceeds the upper bound from the magnetized white dwarfs~\cite{mwd}, hence this model is not viable. 

\

\noindent \cite{carenzamarsh} reconsiders the ALP scenarios discussed in the present  paper. In particular, it focuses the attention on the case in which the photon-to-ALP conversion takes place in the host galaxy, supposed to be identical to the Milky Way. It assumes again that the  power-law flux observed by {\it Fermi/LAT} for $100 \, {\rm MeV} \lesssim {\cal E} \lesssim 1 \, {\rm GeV}$ is extrapolated to the highest considered energies, contrary to the expectation from GRB 190114C. It emphasizes that both LHAASO and Carpet-2 results cannot be explained with the flux in Eq. (\ref{10112022a}), which we however consider unrealistic. Moreover, in order to explain the LHAASO events a value of $g_{a \gamma \gamma}$ in disagreement with the upper bound from the magnetized white dwarfs~\cite{mwd} is needed. Therefore, also this model is not viable.

\

We stress that among the above models the one with the smallest ALP mass $m_{\rm ALP} \simeq 10^{- 10} \, {\rm eV}$ is~\cite{grtClu}, hence this ALP is a good candidate for the dark matter~\cite{arias}.

\

Some remarks are now in order. Our main assumptions are as follows.

\begin{enumerate} 

\item Both the LHAASO and the Carpet-2 collaborations have {\it really} observed the GRB 221009A. We note that the latter result has been called into question by some authors: for instance~\cite{bhm} states that: `{\it a more simple explanation would be a misidentification of a charged cosmic-ray air shower}'. Another question is: why has the Carpet-2 event not been detected also by LHAASO? A possibile answer might be that such an event was not anymore in the field of view of LHAASO at 4536 s after $t_0$ because of the rotation of the Earth. Another possibility is that some selection cut prevented such a detection. The LHAASO collaboration should clarify these points.

\item The EBL {\it prevents} the observability of both the LHAASO and the Carpet-2 observations. This reflects a widespread belief, which is also reported in~\cite{carpet}: `{\it high-energy photons attenuate through production of electron-positron pairs on cosmic background radiation, and $250 \, {\rm TeV}$ photons (as well as $18 \, {\rm TeV}$ photons detected by LHAASO) cannot reach us from the assumed GRB redshift
$z = 0.151$ unless unconventional particle physics is involved. Examples are
axion-like particles}'. Obviously, this depends on the unknown emitted flux, and some authors have questioned the need of new physics (see e.g.~\cite{zhao,zhang}). We believe that this issue will be settled when the LHAASO and Carpet-2 collaborations will release their spectra. Another source of uncertainty is the level of the EBL, which varies depending on the considered paper: some values of $\tau_{\rm CP} (18 \, {\rm TeV})$ are reported in Table 1 of~\cite{bhm} but the EBL model considered in this paper~\cite{franceschinirodighiero} is not included, which is the most recent reference apart from~\cite{sl2021} (note that~\cite{sl2021} has $\tau_{\rm CP} (18 \, {\rm TeV}) \simeq 19.1$ while we take more conservatively $\tau^{\rm FR}_{\rm CP} (18 \, {\rm TeV}) \simeq 14$).

\end{enumerate} 

Moreover, we have stressed that the extrapolation of Eq. (\ref{10112022a}) to arbitrarily high energies has no physical motivation, and -- in analogy with the case of GRB 190114C -- we expect that the inverse Compton peak should be present in the observed SED of GRB 221009A. This is our main message, which remains true even if the ALP scenarios will turn out to be unwarranted. Anyway, only the knowledge of the spectra observed by the LHAASO and Carpet-2 collaborations will shed light on the nature of GRB 221009A in general, and on the need of ALP scenarios in particular.

\section*{Acknowledgments}

We thank Gabriele Ghisellini and Giovanni Pareschi for very useful discussions. The work of G. G. is supported by a contribution from the grant ASI-INAF 2015-023-R.1. The work of M. R. is supported by an INFN TAsP grant. F. T. acknowledge contribution from the grant INAF CTA--SKA, `Probing particle acceleration and $\gamma$-ray propagation with CTA and its precursors' and the INAF Main Stream project `High-energy extragalactic astrophysics: toward the Cherenkov Telescope Array'.

\end{document}